\documentclass[11pt]{article}

\newcommand{\beq}{\begin{equation}} 
\newcommand{\eeq}{\end{equation}}
\newcommand{\p}{\partial} 
\newcommand{\bfe}{\mbox{\bf e}}

\newcommand{\no}{\nonumber} 
\renewcommand{\H}{{\mathcal H}}
\newcommand{\Lc}{{\mathcal L}}

\begin{document}

\vspace*{1.in}

\begin{center}
 
                           {\Large {\bf Gravity and the Electroweak Theory }}\\

\vspace{.5in}
                                   {\large {\bf Kenneth Dalton}}  \\
\vspace{.3in} 
                                    e-mail:  kxdalton@yahoo.com \\

\vspace{1.in}

                                         {\bf Abstract}         \\
\end{center}

\vspace{.25in}

        This work shows that gravity creates inertia for the electron.  
        The Lagrangian directly couples the gravitational field with the 
        electron spinor field.  It does so via the covariant spinor
        derivative.  The Lagrangian is invariant with respect to the
        electroweak gauge group $ U(1) \otimes SU(2)_L $.
        
        The field equations are solved for an electron in uniform motion.
        The solution is the Dirac spinor for a massive electron.  In effect,
        gravitational coupling takes the place of the Dirac mass term.

\clearpage

\section*{\large {\bf 1. Metrical Gravity.}}

The formalism of time-dependent, three-dimensional geometry was derived in [1] and 
reviewed in [2].  A brief summary follows.  We introduce a scalar, 3-vector basis
$ e_\mu = (e_0, {\bfe}_i ) $ and define inner products 

\beq
     g_{\mu\nu} = e_\mu \cdot e_\nu = \left( \begin{array}{cccc}
                                                           g_{00}&0&0&0 \\
                                                           0&&&  \\
                                                           0&&g_{ij}&  \\
                                                           0&&&
                                                           \end{array}
                                                            \right)
\eeq
The basis $ e_\mu $ changes from point to point in the manifold
according to the formula

\beq
        \nabla_{\nu} e_{\mu} = e_{\lambda} Q^{\lambda}_{\mu\nu}
\eeq
This separates into scalar and 3-vector parts

\begin{eqnarray}
      \nabla_\nu e_0 & = & e_0 Q^0_{0 \nu} \\
      \nabla_\nu {\bfe}_i & = & {\bfe}_j Q^j_{i \nu}
\end{eqnarray}
By definition  $ Q^0_{j \nu} = Q^i_{0 \nu} \equiv  0. $ The 18 coefficients 
$ Q^i_{jk} $ would suffice in the case of ordinary static, three-dimensional
geometry.  Here, the geometry will be time-dependent, in general, and clock
rates may differ from point to point.  Hence, the additional coefficients
$ Q^i_{j0} $ and $ Q^0_{0\nu}, $ respectively.  All 28 independent 
$Q^{\mu}_{\nu\lambda} $ are derivable from the metric

\begin{eqnarray}
     Q^0_{0\nu} & = & \Gamma^0_{0\nu} = \frac{1}{2} g^{00} \p_\nu g_{00}  \\
     Q^i_{j0} & = & \Gamma^i_{j0} = \frac{1}{2} g^{il} \p_0 g_{lj}  \\
     Q^i_{jk} & = &\Gamma^i_{jk} = \frac{1}{2} g^{il} \left(\p_k g_{jl} + \p_j g_{lk} 
                                                 - \p_l g_{jk}\right)
\end{eqnarray}
where

\beq
   \Gamma^\mu_{\nu\lambda} = \frac{1}{2} g^{\mu\rho}\left(\p_\lambda g_{\nu\rho}
                               + \p_\nu g_{\rho\lambda} - \p_\rho g_{\nu\lambda}\right)
\eeq
are the Christofel coefficients.  The symbols $ \Gamma^\mu_{\nu\lambda} $ are symmetric 
in $\nu\lambda $, while the $ Q^\mu_{\nu\lambda} $ are not.  The following formula
holds good 

\beq
      Q^\mu_{\nu\lambda} = \Gamma^\mu_{\nu\lambda} +
                         g^{\mu\rho}  g_{\lambda\eta}  Q^\eta_{[\nu\rho]}
\eeq
where 

\beq
        Q^\mu_{[\nu\lambda]} \equiv  Q^\mu_{\nu\lambda} - Q^\mu_{\lambda\nu}
\eeq

The gravitational field is introduced by means of the Einstein-Hilbert action

\beq
        \frac{\kappa}{2} \int R \sqrt{-g} \, d^4 x = 
         \frac{\kappa}{2} \int g^{\mu\nu} R_{\mu\nu} \sqrt{-g} \, d^4 x  
\eeq
where $ \kappa = c^4/8\pi G $, and the Ricci tensor is 

\beq
  R_{\mu\nu} = \p_\nu \Gamma^\lambda_{\mu\lambda} - 
                         \p_\lambda \Gamma^\lambda_{\mu\nu}
                         + \Gamma^\lambda_{\rho\nu} \Gamma^\rho_{\mu\lambda}
                         - \Gamma^\lambda_{\lambda\rho}\Gamma^\rho_{\mu\nu}
\eeq
Variation of the gravitational action gives [3]

\beq
     \delta  \int \frac{\kappa}{2}\, g^{\mu\nu} R_{\mu\nu} \sqrt{-g} \, d^4 x
       = \frac{\kappa}{2} \int \left(R_{\mu\nu} - \frac{ 1}{2} g_{\mu\nu} R \right) 
       \delta g^{\mu\nu} \sqrt{-g} \, d^4 x
\eeq
The source of gravitation is expressed in terms 
of a matter Lagrangian  $ \Lc_m $. Variation 
with respect to $ \delta g^{\mu\nu} $ defines the energy tensor

\beq
       \delta \int \Lc_m \, d^4 x = \frac{1}{2} \int T_{\mu\nu} 
                                     \delta g^{\mu\nu} \sqrt{-g} \, d^4 x 
\eeq
Setting the sum of (13) and (14) equal to zero, we obtain the gravitational field
equations

\beq
         \kappa \left(R_{\mu\nu} - \frac{1}{2} g_{\mu\nu} R \right) + T_{\mu\nu} = 0
\eeq
There are seven field equations corresponding to the seven variations $\delta g^{00}$
and $ \delta g^{ij}$.  

The Lagrangian for gravitation may be derived by partial integration of 
the action [3]

\begin{eqnarray}
  \hspace{-.2in}  \frac{\kappa}{2} \int g^{\mu\nu} R_{\mu\nu} \sqrt{-g} \, d^4 x & = &     
 \frac{\kappa}{2} \int \biggl\{ \p_\nu \left( \sqrt{-g} g^{\mu\nu} \Gamma^\lambda_{\mu\lambda}\right)  
     - \p_\lambda \left(\sqrt{-g} g^{\mu\nu} \Gamma^\lambda_{\mu\nu} \right)  \nonumber \\
   &&\hspace{.3in}  + \p_\lambda \left(\sqrt{-g} g^{\mu\nu} \right)\Gamma^\lambda_{\mu\nu}
     - \p_\nu \left( \sqrt{-g} g^{\mu\nu} \right) \Gamma^\lambda_{\mu\lambda} \nonumber \\
 && \hspace{.3in}  + g^{\mu\nu} \left( \Gamma^\lambda_{\rho\nu}\Gamma^\rho_{\mu\lambda}
   - \Gamma^\lambda_{\lambda\rho}  \Gamma^\rho_{\mu\nu} \right)   \sqrt{-g} \, \biggr\} \, d^4 x 
\end{eqnarray}
The first two terms, when converted to boundary integrals, do not contribute to the 
field equations.  With some rearranging, this leaves

\beq
    \frac{\kappa}{2} \int g^{\mu\nu} R_{\mu\nu} \sqrt{-g} \, d^4 x  =   
    \frac{\kappa}{2} \int g^{\mu\nu} \left( \Gamma^\lambda_{\mu\nu}\Gamma^\rho_{\rho\lambda}
   - \Gamma^\lambda_{\rho\nu}  \Gamma^\rho_{\mu\lambda} \right) \sqrt{-g}   \, d^4 x
\eeq
The Lagrangian 

\beq
    \Lc_g = \frac{\kappa}{2}\, g^{\mu\nu} \left( 
\Gamma^\lambda_{\mu\nu}\Gamma^\rho_{\rho\lambda}
           - \Gamma^\lambda_{\rho\nu}  \Gamma^\rho_{\mu\lambda} \right) \sqrt{-g}
\eeq
depends upon the metric $ g^{\mu\nu} $ and its first derivatives $ \p_\lambda g^{\mu\nu} $.
Therefore, variation (13) now takes the form 

\begin{eqnarray}
  \delta \int \Lc_g \, d^4 x & = & \int \biggl[  \frac{\p \Lc_g}{\p 
g^{\mu\nu} }\delta g^{\mu\nu}
      + \frac{\p\Lc_g}{\p(\p_\lambda g^{\mu\nu})} \delta (\p_\lambda 
              g^{\mu\nu}) \biggr]\, d^4 x \nonumber \\
     & = & \int \biggl[  \frac{\p \Lc_g}{\p g^{\mu\nu} }
      - \p_\lambda \frac{\p\Lc_g}{\p(\p_\lambda g^{\mu\nu})} \biggr] 
\delta g^{\mu\nu} d^4 x
\end{eqnarray}
Variation of the matter action (14) takes a similar form 

\beq
  \delta \int \Lc_m \, d^4 x = \int \biggl[  \frac{\p \Lc_m}{\p 
g^{\mu\nu} }
      - \p_\lambda \frac{\p\Lc_m}{\p(\p_\lambda g^{\mu\nu})} \biggr] 
\delta g^{\mu\nu} d^4 x
\eeq
Defining the total Lagrangian

\beq
      \Lc = \Lc_g + \Lc_m 
\eeq
and setting the sum of (19) and (20) equal to zero, we obtain the gravitational field
equations 

\beq
    \frac{\p \Lc}{\p g^{\mu\nu} } - \p_\lambda \frac{\p\Lc}{\p(\p_\lambda g^{\mu\nu})} = 0 
\eeq
The Lagrange equations have a direct bearing upon the question of energy, which is defined
in terms of $ \Lc $ (appendix C).

\clearpage

\section*{\large {\bf 2. Tetrads. }}

In order to accommodate spinors, the system $ e_\mu $ can be expanded in terms of tetrads
on a Pauli basis 

\beq
     e_\mu  = e^\alpha_{\,\,\,\mu}(x)\, \sigma_\alpha
\eeq
where

\beq
       e^\alpha_{\,\,\, \mu} = \left( \begin{array}{cccc}
                                                            e^0_{\,\,\,0}&0&0&0 \\
                                                            0&&&  \\
                                                            0&&e^a_{\,\,\,i}&  \\
                                                            0&&&
                                                            \end{array}
                                                             \right)  \hspace{.3in}
\eeq
The tetrads can be chosen such that

\beq
    e^1_{\,\,\,2} = e^2_{\,\,\,1} \;\;\;\; e^2_{\,\,\,3} = e^3_{\,\,\,2} \;\;\;\;
        e^3_{\,\,\,1} = e^1_{\,\,\,3}
\eeq
leaving seven independent functions $ e^\alpha_{\,\,\, \mu}(x) $.
The metric now depends upon the underlying tetrad field

\beq
  g_{\mu\nu}   = \eta_{\alpha\beta} \, e^\alpha_{\,\,\,\mu} e^\beta_{\,\,\,\nu}
\eeq

As shown in appendix A, the covariant spinor derivative 

\beq
         D_\mu \psi = \p_\mu \psi + iqA_\mu \psi + \Gamma_\mu \psi
\eeq
yields the Lagrangian (adding the kinetic term for $ A_\mu $)

\begin{eqnarray}
  L_e & = & \frac{i}{2}\hbar c\bigl[\overline{\psi} \gamma^\mu \p_\mu \psi 
        - (\p_\mu \overline{\psi})\gamma^\mu \psi \bigr]  
        + \frac{\hbar c}{4} \overline{\psi} \gamma_5 \hat{\gamma}_\delta \psi
          \, \epsilon^{\delta\alpha\beta\gamma} e_\alpha^{\,\,\,\lambda} 
             e_\gamma^{\,\,\,\nu} \p_\nu e_{\beta\lambda} \no \\
      & &  - \hbar c q\overline{\psi} \gamma^\mu \psi A_\mu     
                         -\frac{1}{4} F_{\mu\nu} F^{\mu\nu} 
\end{eqnarray}
$\hat{\gamma}^\alpha $ are the constant Dirac matrices, while 
$  \gamma^\mu (x) = e_\alpha^{\,\,\,\mu} (x)\, \hat{\gamma}^\alpha $.
The electron field equation is found by varying $ \overline{\psi} $

\beq
   \frac{\p \Lc_e}{\p \overline{\psi}} 
                 - \p_\mu \frac{\p \Lc_e}{\p (\p_\mu 
\overline{\psi})} = 0
\eeq
\newpage \noindent
Substitution of 

\begin{eqnarray}
  \frac{\p\Lc_e}{\p \overline{\psi}} & =  & 
        \hbar c\left\{ \frac{i}{2} \gamma^\mu \p_\mu \psi
       - q \gamma^\mu \psi A_\mu + \frac{1}{4} \gamma_5 \hat{\gamma}_\delta \psi
      \, \epsilon^{\delta\alpha\beta\gamma} e_\alpha^{\,\,\,\lambda} e_\gamma^{\,\,\,\nu}
        \p_\nu e_{\beta\lambda} \right\} \sqrt{-g}              \no     \\ 
        \\            
           \frac{\p \Lc_e}{\p(\p_\mu \overline{\psi})}
                           &  = & -\frac{i}{2} \hbar c \sqrt{-g} \,\gamma^\mu \psi 
\end{eqnarray}
gives 

\begin{eqnarray}
        & &  i \gamma^\mu\p_\mu \psi + 
    \frac{i}{2} \frac{1}{\sqrt{-g}} \p_\mu ( \sqrt{-g} \,\gamma^\mu )\psi
             - q\gamma^\mu \psi A_\mu   \no \\
        & &  \hspace{1.3in} +\frac{1}{4} \gamma_5 \hat{\gamma}_\delta \psi 
      \, \epsilon^{\delta\alpha\beta\gamma}e_\alpha^{\,\,\,\lambda} 
       e_\gamma^{\,\,\,\nu} \p_\nu e_{\beta\lambda} = 0
\end{eqnarray}
A similar calculation with respect to $\delta \psi $ yields the conjugate equation 

\begin{eqnarray}
  & &  -i (\p_\mu \overline{\psi}) \gamma^\mu 
    - \frac{i}{2} \overline{\psi}\frac{1}{\sqrt{-g}} \p_\mu (\sqrt{-g} \,\gamma^\mu) 
     - q \overline{\psi} \gamma^\mu A_\mu    \no \\
  & & \hspace{1.3in}   + \frac{1}{4} \overline{\psi} \gamma_5 \hat{\gamma}_\delta 
   \, \epsilon^{\delta\alpha\beta\gamma} e_\alpha^{\,\,\,\lambda} 
          e_\gamma^{\,\,\,\nu} \p_\nu e_{\beta\lambda} = 0 
\end{eqnarray}
Multiply the first equation by $ \overline{\psi}  $ the second by $ \psi $ 
and subtract, to find conservation of charge

\beq
         \p_\mu \left(\sqrt{-g} \, \overline{\psi}\, \gamma^\mu \psi \right) = 0
\eeq

\vspace{.7in}

In the following sections, we will be concerned with oscillating gravitational fields
of very small amplitude.  The coordinate system is taken to be nearly rectangular

\begin{eqnarray}
            g_{\mu\nu} & = & \eta_{\mu\nu} + h_{\mu\nu} \\
            g^{\mu\nu} & = & \eta^{\mu\nu} - h^{\mu\nu}  
\end{eqnarray}
\newpage
\noindent
The indices of $ h_{\mu\nu}  $ are raised by $ \eta^{\mu\nu} $; for example, in lowest
order 

\begin{eqnarray}
  g_{\mu\lambda} \, g^{\lambda\nu} & = & (\eta_{\mu\lambda} + h_{\mu\lambda})
                                      (\eta^{\lambda\nu} - h^{\lambda\nu}) \no \\
                                & = & \delta^\nu_\mu + h^\nu_\mu - h^\nu_\mu = \delta^\nu_\mu 
\end{eqnarray}
The tetrads are expressed in a similar fashion
\begin{eqnarray}
             e^\alpha_{\,\,\,\mu} & = & \delta^\alpha_{\,\,\,\mu} + \xi^\alpha_{\,\,\,\mu}  \\
             e_\alpha^{\,\,\,\mu} & = & \delta_\alpha^{\,\,\,\mu} - \xi_\alpha^{\,\,\,\mu}
\end{eqnarray}
so that

\begin{eqnarray}
      e^\alpha_{\,\,\,\mu}\, e_\alpha^{\,\,\,\nu} & = & \delta^\nu_\mu 
             =  (\delta^\alpha_{\,\,\, \mu} + \xi^\alpha_{\,\,\, \mu})
                             (\delta_\alpha^{\,\,\, \nu} - \xi_\alpha^{\,\,\, \nu})  \no \\
                   & = & \delta^\nu_\mu + \delta_\alpha^{\,\,\, \nu}\xi^\alpha_{\,\,\, \mu}
                                     - \delta^\alpha_{\,\,\, \mu}\xi_\alpha^{\,\,\, \nu}  
\end{eqnarray}
In order to simplify the notation, we mix indices obtaining

\beq
     \xi^\nu_{\,\,\, \mu} = \xi_\mu^{\,\,\, \nu} 
\eeq
In terms of tetrads, the metric is 

\begin{eqnarray}
  g_{\mu\nu} & = & \eta_{\alpha\beta} \, e^\alpha_{\,\,\,\mu} e^\beta_{\,\,\,\nu} \no \\
    & = &   \eta_{\mu\nu} + \eta_{\alpha\beta}(\delta^\alpha_{\,\,\, \mu}\xi^\beta_{\,\,\, \nu}
                                         + \delta^\alpha_{\,\,\, \nu}\xi^\beta_{\,\,\, \mu})
\end{eqnarray}
so that 

\begin{eqnarray}
       h_{\mu\nu} & = &\eta_{\alpha\beta}
                     (\delta^\alpha_{\,\,\, \mu}\xi^\beta_{\,\,\, \nu}
                     + \delta^\alpha_{\,\,\, \nu}\xi^\beta_{\,\,\, \mu}) \no \\
                  & = & 2\xi_{\mu\nu}   
\end{eqnarray}

\clearpage
\section*{\large {\bf 3. \boldmath{$ U(1) \otimes SU(2)_L$ } Gauge Invariance. }
                   \rm[4, 5, 6]}

The gravitational coupling term in $ L_e $ (28) contains the factor 
$ \overline{\psi} \gamma_5 \hat{\gamma}_\delta \psi $.  This factor does not mix
right- and left-handed spinor components, $ \psi_R $ and $ \psi_L $.  In order to
prove this, set

\beq
 \psi = \psi_R + \psi_L = \frac{1 + \gamma_5}{2}\, \psi + 
     \frac{1 - \gamma_5}{2} \,\psi
\eeq
where $ (\gamma_5)^2 = 1 $ and $ \gamma_5 \gamma_\delta = - \gamma_\delta \gamma_5 $.
Also, $ \overline{\psi}_R = \overline{\psi}\,\,\frac{1 - \gamma_5}{2}\, $ and 
$\, \overline{\psi}_L = \overline{\psi}\,\,\frac{1 + \gamma_5}{2} $.\newline
In the expansion

\begin{eqnarray}
\overline{\psi} \gamma_5 \hat{\gamma}_\delta \psi & = &
  (\overline{\psi}_R + \overline{\psi}_L )\, \gamma_5 \gamma_\delta \,
(\psi_R + \psi_L) \no \\
  & = & \overline{\psi}_R \gamma_5 \gamma_\delta \psi_R 
         + \overline{\psi}_L \gamma_5 \gamma_\delta \psi_L 
         + \overline{\psi}_R \gamma_5 \gamma_\delta \psi_L 
         + \overline{\psi}_L \gamma_5 \gamma_\delta \psi_R 
\end{eqnarray}
the mixed terms are identically zero.  For example,

\beq
  \overline{\psi}_R \gamma_5 \gamma_\delta \psi_L =  
   \overline{\psi}\,\frac{1-\gamma_5}{2} \,\gamma_5 \gamma_\delta 
           \,\frac{1-\gamma_5}{2}\,\psi
  = \overline{\psi}\,\gamma_5 \gamma_\delta 
           \,\frac{1-(\gamma_5)^2}{4} \,\psi = 0
\eeq
Therefore, 

\beq
  \overline{\psi} \gamma_5 \hat{\gamma}_\delta \psi = 
    \overline{\psi}_R \gamma_5 \gamma_\delta \psi_R 
         + \overline{\psi}_L \gamma_5 \gamma_\delta \psi_L
\eeq

An expression of this type will be invariant under 
$ U(1) \otimes SU(2)_L $ gauge transformations.
\footnote{The Dirac mass term $\, m \overline{\psi} \psi =
m(\overline{\psi}_R {\psi}_L + \overline{\psi}_L {\psi}_R)\, $
mixes right- and left-handed spinors and cannot appear in
the electroweak Lagrangian.}
Introduce the right-handed singlet $ \psi_R = e_R $ and
left-handed doublet $\psi_L = \left(\begin{array}{c} \nu_L \\
e_L \end{array}\right) $  in order to form the Lagrangian

\begin{eqnarray}
 L_{e-w} &=&  \frac{i}{2}\hbar c\bigl[\overline{\psi}_R \gamma^\mu \p_\mu \psi_R 
    + \overline{\psi}_L \gamma^\mu \p_\mu \psi_L \bigr] + \mbox{\rm h.c.} \no \\
  & + & \frac{\hbar c}{4} \bigl[\overline{\psi}_R \gamma_5 \hat{\gamma}_\delta \psi_R
         +\overline{\psi}_L \gamma_5 \hat{\gamma}_\delta \psi_L \bigr]
       \, \epsilon^{\delta\alpha\beta\gamma} e_\alpha^{\,\,\,\lambda} 
             e_\gamma^{\,\,\,\nu} \p_\nu e_{\beta\lambda} 
  + L_{int}
\end{eqnarray}
$ L_{int} $ contains the electroweak interaction terms as well as kinetic
terms for $ A_\mu, W^{\pm}_\mu $, and $ Z^0_\mu $.

\clearpage
\section*{\large {\bf 4.  An Electron in Uniform Motion. }}

In the previous article [2], the gravitational field was found for a
spin up electron at rest

\beq
            \psi = \frac{1}{\sqrt{V}} \left( \begin{array}{c} 
                                                            1  \\
                                                            0  \\
                                                            0  \\
                                                            0
                                              \end{array}
                                 \right) \mbox{\rm exp}\,(-i\omega t)
\eeq
which took the form 

\beq
        e^\alpha_{\,\,\, \mu} = \left( \begin{array}{cccc}
                                          1&0&0&0 \\
                                          0&e^1_{\,\,\,1}&e^1_{\,\,\,2}&0  \\
                                          0&e^2_{\,\,\,1}&e^2_{\,\,\,2}&0  \\
                                          0&0&0&1
                                        \end{array} \right)
\eeq                                                            
Here, we seek the solution for an electron in uniform motion along $x^3$

\beq
      \psi = \frac{N}{\sqrt{V}} \left( \begin{array}{c} 
                                         u_1  \\
                                         u_2  \\
                                         u_3  \\
                                         u_4
                                        \end{array}
     \right) \mbox{\rm exp}\,(-ik_\mu x^\mu) \hspace{.3in} k_\mu =  (k_0,k_3)
\eeq
The second term in the electron equation (32) is zero,
\footnote{The trial solution (51) gives $ \p_\mu \psi = -ik_\mu \psi $ and
$ \p_\mu \overline{\psi} = ik_\mu \overline{\psi}. $ From the conservation 
law (34), it follows that $ \overline{\psi}\, \p_\mu (\sqrt{-g}\, \gamma^\mu )\psi = 0. $}
which leaves

\beq
   i \gamma^\mu\p_\mu \psi + \frac{1}{4} \gamma_5 \hat{\gamma}_\delta \psi 
     \, \epsilon^{\delta\alpha\beta\gamma}e_\alpha^{\,\,\,\lambda} 
       e_\gamma^{\,\,\,\nu} \p_\nu e_{\beta\lambda} = 0
\eeq
In the present case, 

\beq
   \p_0 \psi = -ik_0 \psi \,\,\,\,\mbox{{\rm and}}\,\,\,\,\p_3\psi = -ik_3\psi
\eeq
so that

\beq
  (k_0 \hat{\gamma}^0 + k_3  \hat{\gamma}^3) \psi
         +  \frac{1}{4}\gamma_5\hat{\gamma}_3 \psi \,\epsilon^{3ab0}              
                    e_a^{\,\,\,l} \p_0 e_{bl}
         +  \frac{1}{4}\gamma_5\hat{\gamma}_0 \psi \,\epsilon^{0ab3} 
                    e_a^{\,\,\,l} \p_3 e_{bl}  = 0
\eeq
where the tetrad assumes the form (50).  The Dirac matrix representation is

\beq
         \hat{\gamma}^0 = \left( \begin{array}{cc}
                                                     \sigma_0 &0 \\
                                                      0& -\sigma_0
                                               \end{array}
                                                \right) 
   \;\;\;\; \hat{\gamma}^a = \left( \begin{array}{cc}
                                                     0& \sigma_a \\
                                                    -\sigma_a &0 
                                                \end{array}
                                                   \right)  
   \;\;\;\;  \gamma_5 = \left( \begin{array}{cc}
                                           0 & 1  \\
                                           1 & 0 
                                          \end{array}
                                              \right)
\eeq
while $ \epsilon^{0123} = -1 $.  Substitution yields four equations

\begin{eqnarray}
   k_0u_1 + k_3u_3 + \frac{1}{4}u_1(e_1^{\,\,\,n}\p_0 e_{2n}
                       - e_2^{\,\,\,n}\p_0e_{1n})
                               + \frac{1}{4}u_3(e_1^{\,\,\,n}\p_3 e_{2n} 
                       - e_2^{\,\,\,n}\p_3 e_{1n})=0 &&\no  \\
                  \\
  k_0u_2 - k_3u_4 - \frac{1}{4}u_2(e_1^{\,\,\,n}\p_0 e_{2n}
                       - e_2^{\,\,\,n}\p_0e_{1n})
                   + \frac{1}{4}u_4(e_1^{\,\,\,n}\p_3 e_{2n} 
                       - e_2^{\,\,\,n}\p_3 e_{1n})=0 && \no\\ 
        \\
   k_3u_1 + k_0u_3 + \frac{1}{4}u_3(e_1^{\,\,\,n}\p_0 e_{2n}
                       - e_2^{\,\,\,n}\p_0e_{1n})
          + \frac{1}{4}u_1(e_1^{\,\,\,n}\p_3 e_{2n} 
                       - e_2^{\,\,\,n}\p_3 e_{1n})=0 &&\no  \\          
      \\ 
  k_3u_2 - k_0u_4 + \frac{1}{4}u_4(e_1^{\,\,\,n}\p_0 e_{2n}
                       - e_2^{\,\,\,n}\p_0e_{1n})
             - \frac{1}{4}u_2(e_1^{\,\,\,n}\p_3 e_{2n} 
                       - e_2^{\,\,\,n}\p_3 e_{1n})=0  &&\no\\
\end{eqnarray}
These equations pair $ (u_1,u_3) $ or $ (u_2,u_4) $.
Consider the case $ u_2 = u_4 = 0 $: the equations will be satisfied 
for all values of $ u_1 $ and $ u_3 $, if the coefficients are zero 
\footnote{For the case $ u_1 = u_3 = 0 $, the coefficients of $ u _2 $ 
and $ u_4 $ must be zero

\begin{eqnarray}
     4k_0 - e_1^{\,\,\,n} \p_0e_{2n} + e_2^{\,\,\,n} \p_0e_{1n} & = & 0 \no \\
     4k_3 - e_1^{\,\,\,n} \p_3e_{2n} + e_2^{\,\,\,n} \p_3e_{1n} & = & 0 \no
\end{eqnarray}
Thus, the two cases are mutually exclusive.}

\begin{eqnarray}
     4k_0 + e_1^{\,\,\,n} \p_0e_{2n} - e_2^{\,\,\,n} \p_0e_{1n} & = & 0 \\
     4k_3 + e_1^{\,\,\,n} \p_3e_{2n} - e_2^{\,\,\,n} \p_3e_{1n} & = & 0
\end{eqnarray}
When expressed in terms of the $ \xi^\alpha_{\,\,\,\mu} $(38), these
equations read

\begin{eqnarray}
     4k_0 + (\xi^1_{\,\,\,1} - \xi^2_{\,\,\,2})\p_0 \xi^1_{\,\,\,2}
          -  \xi^1_{\,\,\,2}\p_0 (\xi^1_{\,\,\,1} - \xi^2_{\,\,\,2}) & = & 0 \\
     4k_3 + (\xi^1_{\,\,\,1} - \xi^2_{\,\,\,2})\p_3 \xi^1_{\,\,\,2}
          -  \xi^1_{\,\,\,2}\p_3 (\xi^1_{\,\,\,1} - \xi^2_{\,\,\,2}) & = & 0 
\end{eqnarray}
The first order terms vanish, leaving only second order terms. 
A solution is given by 

\begin{eqnarray}
      (\xi^1_{\,\,\,1} - \xi^2_{\,\,\,2}) &=& 2a\, \mbox{\rm cos}(-k'_\mu x^\mu)\\
        \xi^1_{\,\,\,2} &=& a \,\mbox{\rm sin}(-k'_\mu x^\mu)
\end{eqnarray}
where the amplitude $a$ is small compared with 1.  Substitution yields 

\begin{eqnarray}
           k_0 &=& \frac{1}{2}a^2 k'_0 \ll k'_0 \\
           k_3 &=& \frac{1}{2}a^2 k'_3 \ll k'_3
\end{eqnarray}
Therefore, the electron's frequency and wave vector are much smaller 
than those of the gravitational field.  Their phase velocities are equal,
$ k_0/k_3 = k'_0/k'_3 $.

The spinor  
\beq
      \psi = \frac{N}{\sqrt{V}} \left( \begin{array}{c} 
                                         u_1  \\
                                         0  \\
                                         u_3  \\
                                         0
                                        \end{array}
     \right) \mbox{\rm exp}\,(-ik_\mu x^\mu) \hspace{.3in} k_\mu =  (k_0,k_3)
\eeq
has the form of a Dirac spinor for a positive energy electron of mass $m$,
moving along $x^3$.  Therefore, it will satisfy the Dirac equation

\beq
   i \gamma^\mu \p_\mu \psi - \frac{mc}{\hbar} \psi = 0 
\eeq
By giving rise to such a solution, gravity creates inertia for the electron.
Components $ u_1 $ and $ u_3 $ now satisfy

\begin{eqnarray}
    k_0 u_1 - k_3 u_3 - \frac{mc}{\hbar}u_1 & = &0 \\
    k_3 u_1 + k_0 u_3 + \frac{mc}{\hbar}u_3 & = &0
\end{eqnarray}
so that 

\beq
     u_3 = \frac{-\hbar k_3 u_1}{(\hbar k_0 + mc)} \,\,\,\, 
        \mbox{\rm and}\,\,\,\,\,\, {\hbar}^2 k_0^{\,\,2} 
         = {\hbar}^2 k_3^{\,\,2} + m^2c^2 
\eeq
The final spinor is 

\beq
      \psi = \frac{N}{\sqrt{V}} \left( \begin{array}{c} 
                                         1  \\
                                         0  \\
                                         \frac{-\hbar k_3 }{(\hbar k_0 + mc)}  \\
                                         0
                                        \end{array}
     \right) \mbox{\rm exp}\,(-ik_\mu x^\mu) \hspace{.3in} k_\mu =  (k_0,k_3)
\eeq
where

\beq
    N^2 = \frac{\hbar k_0 + mc}{2\hbar k_0} 
\eeq

\newpage
We turn now to the gravitational field equations (15), retaining only the 
largest terms in the Ricci tensor

\begin{eqnarray}
   R_{\mu\nu} & \doteq & \p_\nu \Gamma^\lambda_{\mu\lambda} - 
                              \p_\lambda \Gamma^\lambda_{\mu\nu}         \\
      \p_\rho g_{\mu\nu} & = & \eta_{\alpha\beta}
           \left(e^\alpha_{\,\,\,\mu} \p_\rho e^\beta_{\,\,\,\nu} 
             + e^\alpha_{\,\,\,\nu} \p_\rho e^\beta_{\,\,\,\mu} \right)  \no \\
      & \doteq & \eta_{\alpha\beta}
             \left(\delta^\alpha_{\,\,\,\mu} \p_\rho \xi^\beta_{\,\,\,\nu} 
          + \delta^\alpha_{\,\,\,\nu} \p_\rho \xi^\beta_{\,\,\,\mu} \right) \no \\
      & = & 2\, \p_\rho \xi_{\mu\nu}
\end{eqnarray}
It follows that

\beq
  R_{\mu\nu} = \eta^{\lambda\rho}
    \{\p_\lambda\p_\rho \xi_{\mu\nu}  + \p_\mu\p_\nu \xi_{\lambda\rho}
     - \p_\mu\p_\lambda \xi_{\rho\nu} - \p_\nu\p_\lambda \xi_{\rho\mu} \}
\eeq
The field equations are

\begin{eqnarray}
 \frac{\kappa\lambda^3}{V} \p_3\p_3(\xi^1_{\,\,\,1} + \xi^2_{\,\,\,2}) + T_{00}&=&0 \\
 \frac{\kappa\lambda^3}{V} (\p_0\p_0 - \p_3\p_3)\xi^2_{\,\,\,2} + T_{11} &=& 0     \\ 
 \frac{\kappa\lambda^3}{V} (\p_0\p_0 - \p_3\p_3)\xi^1_{\,\,\,1} + T_{22} &=& 0     \\
 \frac{\kappa\lambda^3}{V} \p_0\p_0(\xi^1_{\,\,\,1} + \xi^2_{\,\,\,2}) + T_{33} &=& 0 \\
 \frac{\kappa\lambda^3}{V} (\p_0\p_0 - \p_3\p_3)\xi^1_{\,\,\,2} - T_{12} &=& 0    
\end{eqnarray}
A length parameter $\lambda$ is introduced together with the arbitrary
volume of integration, $V$.  The energy tensor is derived in appendix B (128).
Make use of the spinor (73) and  

\begin{eqnarray}
   (u_1^* u_1 + u_3^* u_3) & = & \frac{2\hbar k_0}{\hbar k_0 + mc} \\
   (u_1^* u_3 + u_3^* u_1) & = & \frac{-2\hbar k_3}{\hbar k_0 + mc} 
\end{eqnarray}
to find 
\newpage
\begin{eqnarray}
  T_{00} & = & \frac{\hbar c k_0}{V} + \frac{\hbar c}{4V}
      (e_1^{\,\,\,n}\p_0 e_{2n} - e_2^{\,\,\,n}\p_0 e_{1n}) \\
  T_{11} & = & - \frac{\hbar c}{2V}\,\p_0 e_{21}  
         + \frac{\hbar ck_3}{2k_0 V}\, \p_3 e_{21}     \\
  T_{22} & = &  \frac{\hbar c}{2V}\, \p_0 e_{12}
       - \frac{\hbar ck_3}{2k_0 V}\, \p_3 e_{12}   \\  
  T_{33} & = & \frac{\hbar ck_3^{\,\,2}}{k_0 V}
       + \frac{\hbar ck_3}{4k_0 V}
      (e_1^{\,\,\,n}\p_3 e_{2n} - e_2^{\,\,\,n}\p_3 e_{1n}) \\
  T_{12} & = & -\frac{\hbar c}{4V}\,\p_0 (e_{11} -  e_{22})  
      + \frac{\hbar ck_3}{4k_0 V}\, \p_3 (e_{11} -  e_{22}) 
\end{eqnarray}
Comparison with (60, 61) shows that $ T_{00} = T_{33} = 0 $; also,
$ T_{11} = - T_{22} $. The field equations then give  
$ \,\xi^1_{\,\,\,1} = - \xi^2_{\,\,\,2} \, $ so that two 
independent equations remain

\begin{eqnarray}
   \kappa\lambda^3 (\p_0\p_0 - \p_3\p_3)\xi^1_{\,\,\,1} 
     - \frac{\hbar c}{2}\, \p_0\xi^1_{\,\,\,2}
     + \frac{\hbar ck_3}{2k_0}\, \p_3\xi^1_{\,\,\,2} = 0 \\
   \kappa\lambda^3 (\p_0\p_0 - \p_3\p_3)\xi^1_{\,\,\,2}
     + \frac{\hbar c}{2}\, \p_0\xi^1_{\,\,\,1}
     - \frac{\hbar ck_3}{2k_0}\, \p_3\xi^1_{\,\,\,1} = 0 
\end{eqnarray}
These equations are satisfied by (64, 65), if 

\beq 
    \kappa\lambda^3 = \frac{\hbar c}{2k'_0}
\eeq
The gravitational field is given by

\begin{eqnarray}
    e^1_{\,\,\,1} &=& 1 + a \,\mbox{\rm cos}(-k'_\mu x^\mu) \\
    e^2_{\,\,\,2} &=& 1 - a \,\mbox{\rm cos}(-k'_\mu x^\mu)  \\
    e^1_{\,\,\,2} &=& e^2_{\,\,\,1} = a \,\mbox{\rm sin}(-k'_\mu x^\mu)
\end{eqnarray}
where $ k'_\mu = (k'_0,k'_3). $ The metrical determinant is constant
$ \sqrt{-g} = 1 - a^2 $ so that the condition 
$ \p_\mu (\sqrt{-g}\,\gamma^\mu) =  0 $ is satisfied (footnote 2).

\newpage

Finally, we calculate the total energy (153, appendix C)

\begin{eqnarray}
   \H & = & \frac{\kappa}{2} \left(
          \eta^{00}  \Gamma^l_{m0}\Gamma^m_{l0} 
            + \eta^{lm}\Gamma^p_{nl}\Gamma^n_{mp} \right)  \no \\
   & -& \frac{i}{2}\hbar c\bigl[\overline{\psi} \gamma^3 \p_3 \psi 
        - (\p_3 \overline{\psi})\gamma^3 \psi \bigr]       
   - \frac{\hbar c}{4} \overline{\psi} \gamma_5 \hat{\gamma}_0 \psi
          \,\epsilon^{0ab3} e_3^{\,\,\,3} e_{am} \p_3 e_b^{\,\,\,m}
\end{eqnarray}       
The last two terms are equal to $ T_{33} = 0 $.  Therefore, the
energy is determined by the gravitational field

\begin{eqnarray}
 \H &=& \frac{\kappa\lambda^3}{V}
    \left\{ (\p_0 \xi^1_{\,\,\,1})^2 + (\p_3 \xi^1_{\,\,\,1})^2
      + (\p_0 \xi^1_{\,\,\,2})^2 + (\p_3 \xi^1_{\,\,\,2})^2 \right\}\no\\
    & = & \frac{\hbar c}{2k'_0 V} a^2 
           \left( k_0^{\prime \,\,2} + k_3^{\prime \,\,2} \right) 
\end{eqnarray} 
Integrate over all of space to find

\beq
     E = \int \H \, d^3x = \hbar \omega + \frac{\hbar c k_3^{\,\,2}}{k_0}
\eeq
A constant energy must be subtracted in order to obtain $ E = \hbar \omega $.

\clearpage

\section*{\large {\bf Appendix A: Electron Lagrangian.}}

The scalar, 3-vector basis changes according to the formula

\beq
        \nabla_\nu e_\mu = e_\lambda Q^\lambda_{\mu\nu} 
\eeq
Expanding $ e_\mu $ in terms of tetrads 

\beq
        e_\mu = e^\alpha_{\,\,\,\mu} \sigma_\alpha
\eeq
we have

\begin{eqnarray}
     \nabla_\nu e_\mu & = & \sigma_\alpha \,\p_\nu e^\alpha_{\,\,\,\mu} 
                    + e^\alpha_{\,\,\,\mu} \nabla_\nu \sigma_\alpha  \no  \\
                        & = & \sigma_\alpha \left(\p_\nu e^\alpha_{\,\,\,\mu} 
                      + e^\beta_{\,\,\,\mu} \omega^\alpha_{\,\,\,\beta\nu} \right)
\end{eqnarray}
where, by definition, 
\footnote{ The $\omega^\alpha_{\,\,\,\beta\nu} $ give the change
of orientation of the orthonormal basis $\sigma_\alpha$ from point to point.  Since 
$\eta_{\alpha\beta} = \frac{1}{2} (\sigma_\alpha\widetilde{\sigma}_\beta 
        + \sigma_\beta\widetilde{\sigma}_\alpha) $  where $ \widetilde{\sigma}_\alpha = 
(\sigma_0, -\sigma_a) $ and $ \nabla_\nu \eta_{\alpha\beta} = \p_\nu \eta_{\alpha\beta}
= 0, $ we have $ \omega_{\alpha\beta\nu} = - \omega_{\beta\alpha\nu} $.  Moreover, 
$ \sigma_0 \sigma_a = \sigma_a $ implies that $ \nabla_\nu \sigma_0 = 0 $ or 
$ \omega_{a0\nu} = 0 $.  This leaves 12 parameters $\omega_{ab\nu}$.
They comprise 3 rotation parameters along each of the four coordinates $x^\nu$.}

\beq
       \nabla_\nu \sigma_\alpha = \sigma_\beta \,\omega^\beta_{\,\,\,\alpha\nu} 
\eeq
Equate the two expressions (99) and (101) to find 

\beq
      e^\alpha_{\,\,\,\lambda} Q^\lambda_{\mu\nu} = \p_\nu e^\alpha_{\,\,\,\mu} 
          + e^\beta_{\,\,\,\mu} \omega^\alpha_{\,\,\,\beta\nu} 
\eeq
Contract this equation with $ e_{\alpha\rho} $ and form the tensor 

\begin{eqnarray}
        g_{\rho\lambda} Q^\lambda_{[\mu\nu]} & = & 
     e_{\alpha\rho}(\p_\nu e^\alpha_{\,\,\,\mu} - \p_\mu e^\alpha_{\,\,\,\nu}) 
    + e_{\alpha\rho}(e^\beta_{\,\,\,\mu}\omega^\alpha_{\,\,\,\beta\nu}
           - e^\beta_{\,\,\,\nu}\omega^\alpha_{\,\,\,\beta\mu})                \no \\
      & = & e^\alpha_{\,\,\,\rho}(\p_\nu e_{\alpha\mu} - \p_\mu e_{\alpha\nu}) 
           + \omega_{\rho\mu\nu} - \omega_{\rho\nu\mu} 
\end{eqnarray}
From (9), it follows that the totally anti-symmetric tensor

\beq
         Q_{[ \mu\nu\lambda ]} = g_{\mu\rho} Q^\rho_{[ \nu\lambda ]} 
        + g_{\nu\rho} Q^\rho_{[ \lambda\mu ]} + g_{\lambda\rho} Q^\rho_{[ \mu\nu ]} = 0
\eeq
Therefore,

\beq
         0 = e^\alpha_{\,\,[ \rho} \p_\nu e_{\alpha\mu ]} + \omega_{ [ \rho\mu\nu ]}
\eeq
\newpage
\noindent
or

\beq
       \omega_{ [\mu\nu\lambda] } = e^\alpha_{\,\,[ \mu} \p_\nu e_{\alpha\lambda ]}
\eeq
where

\beq
  \omega_{[\mu\nu\lambda]} \equiv \frac{1}{6} \biggl(\omega_{\mu\nu\lambda} 
       + \omega_{\nu\lambda\mu} + \omega_{\lambda\mu\nu}
       - \omega_{\nu\mu\lambda} - \omega_{\mu\lambda\nu}
       - \omega_{\lambda\nu\mu} \biggr)
\eeq

\indent
The covariant spinor derivative is [7, 8] 

\beq
      D_\mu \psi = \p_\mu \psi + iqA_\mu \psi + \Gamma_\mu \psi
\eeq
where

\beq
 \Gamma_\mu = \frac{1}{8} \left(\hat{\gamma}^\alpha \hat{\gamma}^\beta 
    - \hat{\gamma}^\beta \hat{\gamma}^\alpha \right) \omega_{\alpha\beta\mu} 
  = \frac{1}{4} \hat{\gamma}^{[\alpha} \hat{\gamma}^{\beta ]} \omega_{\alpha\beta\mu}
\eeq
We have included the U(1) term $iqA_\mu \psi$.  The conjugate 
expression is  

\beq
     D_\mu\overline{\psi} = \p_\mu \overline{\psi} - i qA_\mu 
\overline{\psi} 
              - \overline{\psi} \Gamma_\mu 
\eeq
giving the electron Lagrangian  

\begin{eqnarray} 
    L_e & = & \frac{i}{2} \hbar c \bigl[ \overline{\psi} \gamma^\mu D_\mu \psi 
             - (D_\mu \overline{\psi}) \gamma^\mu \psi \bigr]    \no  \\
      & = &  \frac{i}{2} \hbar c \bigl[ \overline{\psi} \gamma^\mu \p_\mu \psi 
             - (\p_\mu \overline{\psi}) \gamma^\mu \psi \bigr] 
            - \hbar c q\overline{\psi} \gamma^\mu \psi A_\mu   \no  \\
      & & + \, \frac{i}{2} \hbar c \overline{\psi}\left(\gamma^\mu \Gamma_\mu 
               + \Gamma_\mu \gamma^\mu \right) \psi 
\end{eqnarray}
The gravitational coupling term can be expressed in terms of the tetrad 
field as follows:

\begin{eqnarray}
   \gamma^\mu \Gamma_\mu  + \Gamma_\mu \gamma^\mu & = & 
       \frac{1}{4} e_\gamma^{\,\,\,\mu} \left(\hat{\gamma}^\gamma \hat{\gamma}^{[\alpha}
      \hat{\gamma}^{\beta ]} +  \hat{\gamma}^{[\alpha} \hat{\gamma}^{\beta ]} 
          \hat{\gamma}^\gamma \right) \omega_{\alpha\beta\mu}  \no  \\
  & = & \frac{1}{2} \hat{\gamma}^{[\alpha} \hat{\gamma}^\beta 
           \hat{\gamma}^{\gamma ]} \omega_{\alpha\beta\gamma}
\end{eqnarray}
\clearpage
\noindent
where we have used the identity [8]

\beq
        \hat{\gamma}^{[\gamma} \hat{\gamma}^\alpha \hat{\gamma}^{\beta ]}  
            \equiv  \frac{1}{2}  \left( \hat{\gamma}^\gamma \hat{\gamma}^{[\alpha}
            \hat{\gamma}^{\beta ]} +  \hat{\gamma}^{[\alpha} \hat{\gamma}^{\beta ]} 
                    \hat{\gamma}^\gamma \right)
\eeq
The identity [8] 

\beq
        \hat{\gamma}^{[\alpha} \hat{\gamma}^\beta \hat{\gamma}^{\gamma ]}  
     \equiv    -i \gamma_5 \hat{\gamma}_\delta \,\epsilon^{\delta\alpha\beta\gamma}
\eeq
yields
\begin{eqnarray} 
    \gamma^\mu \Gamma_\mu  + \Gamma_\mu \gamma^\mu & = & 
          -\frac{i}{2} \gamma_5 \hat{\gamma}_\delta \,\epsilon^{\delta\alpha\beta\gamma}
            \omega_{\alpha\beta\gamma}                       \no  \\
          & = & -\frac{i}{2} \gamma_5 \hat{\gamma}_\delta \,\epsilon^{\delta\alpha\beta\gamma}
            \omega_{[\alpha\beta\gamma]}
\end{eqnarray}
since $ \epsilon^{\delta\alpha\beta\gamma} $ is totally anti-symmetric.  Substitute
(107) in order to obtain 

\beq
    \gamma^\mu \Gamma_\mu  + \Gamma_\mu \gamma^\mu = 
       -\frac{i}{2} \gamma_5 \hat{\gamma}_\delta \,\epsilon^{\delta\alpha\beta\gamma}
     e_\alpha^{\,\,\,\lambda} e_\gamma^{\,\,\,\nu} \p_\nu e_{\beta\lambda} 
\eeq
and the Lagrangian

\begin{eqnarray}
 L_e & = & \frac{i}{2}\hbar c\bigl[\overline{\psi} \gamma^\mu \p_\mu \psi 
                       - (\p_\mu \overline{\psi})\gamma^\mu \psi \bigr]  
                  - \hbar c q\overline{\psi} \gamma^\mu \psi A_\mu  \no \\
         & &      + \frac{\hbar c}{4} \overline{\psi} \gamma_5 \hat{\gamma}_\delta \psi
               \,\epsilon^{\delta\alpha\beta\gamma} e_\alpha^{\,\,\,\lambda} 
                         e_\gamma^{\,\,\,\nu} \p_\nu e_{\beta\lambda}
\end{eqnarray}

\clearpage

\section*{\large {\bf Appendix B: Electron Energy Tensor.}}

The electron energy tensor is found by varying the tetrad field 
$ e_\alpha^{\,\,\,\mu} $

\begin{eqnarray}
      \delta \int \Lc_e \, d^4 x &=& \int \left(\frac{\p \Lc_e}{\p 
e_\alpha^{\,\,\,\mu} }
       \delta e_\alpha^{\,\,\,\mu}  + \frac{\p \Lc_e}{\p (\p_\lambda 
e_\alpha^{\,\,\,\mu})}
       \delta \p_\lambda e_\alpha^{\,\,\,\mu} \right) \, d^4 x  \no \\
  &=& \int \biggl[ \frac{\p \Lc_e}{ \p e_\alpha^{\,\,\,\mu}} - 
   \p_\lambda \frac{\p \Lc_e}{\p(\p_\lambda e_\alpha^{\,\,\,\mu})} 
\biggr] 
       \delta e_\alpha^{\,\,\,\mu} \, d^4 x   \no \\
 &=& \int \sqrt{-g} \, A_{\mu\nu} e^{\beta\nu} \delta e_\beta^{\,\,\,\mu} \, d^4 x
\end{eqnarray}
where 

\beq
    \sqrt{-g} \, A_{\mu\nu} \equiv  e_{\alpha\nu} 
          \biggl[ \frac{\p \Lc_e}{ \p e_\alpha^{\,\,\mu}} - 
          \p_\lambda \frac{\p \Lc_e}{\p(\p_\lambda e_\alpha^{\,\,\,\mu})} \biggr]
\eeq
The action is invariant under arbitrary rotations of the tetrad [7]. 
An infinitesimal rotation takes the form 

\beq
   \delta e_0^{\,\,\,0} = 0 \; \;\;\;\delta e_a^{\,\,\,i} = 
     \epsilon_a^{\,\,\,b} e_b^{\,\,\,i} \;\;\;\; \epsilon^{ab} = - \epsilon^{ba} 
\eeq
Thus,

\beq
   \delta \int \Lc_e\,d^4 x = \int \sqrt{-g} \, A_{ij} e_a^{\,\,\,i} e_b^{\,\,\,j} 
                                             \epsilon^{ba} \, d^4 x = 0
\eeq
Therefore, the antisymmetric part of $ A_{ij} $ must be zero

\beq
                 \frac{1}{2} (A_{ij} - A_{ji}) = 0
\eeq
and we define the symmetric part to be 

\beq
         T_{\mu\nu } = \frac{1}{2} (A_{\mu\nu} + A_{\nu\mu})
\eeq
It follows that, for an arbitrary variation,

\begin{eqnarray}
   \delta \int \Lc_e \, d^4 x &=& \int \sqrt{-g} \, A_{\mu\nu} 
\frac{1}{2} 
               \left(e^{\beta\mu}\delta e_\beta^{\,\,\,\nu} 
                       + e^{\beta\nu} \delta e_\beta^{\,\,\,\mu}\right)\, d^4 x  \no\\
                  &=& \frac{1}{2} \int  e_{\alpha\nu} 
                        \biggl[ \frac{\p \Lc_e}{ \p e_\alpha^{\,\,\,\mu}} 
                   - \p_\lambda \frac{\p \Lc_e}{\p(\p_\lambda 
e_\alpha^{\,\,\mu})} \biggr]
                            \delta g^{\mu\nu} \, d^4 x  \no \\
               &=& \frac{1}{2} \int T_{\mu\nu }\, \delta g^{\mu\nu} \sqrt{-g} \, d^4 x
\end{eqnarray}
Explicitly,

\begin{eqnarray}
   \frac{\p \Lc_e}{\p e_\alpha^{\,\,\,\mu}} & = & \frac{i}{2} \hbar c
     \bigl[\overline{\psi} \gamma_\lambda \p_\mu \psi - 
     (\p_\mu\overline{\psi})\gamma_\lambda \psi\bigr] e^{\alpha\lambda} \sqrt{-g}
   - \hbar c q\overline{\psi}\gamma_\lambda \psi A_\mu 
     e^{\alpha\lambda}\sqrt{-g} \no\\
 & & \hspace{.3in} + \frac{\hbar c}{4} \overline{\psi} \gamma_5 \hat{\gamma}_\delta \psi
  \, \epsilon^{\delta\alpha\beta\gamma} e_\gamma^{\,\,\,\lambda}
        \left(\p_\lambda e_{\beta\mu} - \p_\mu e_{\beta\lambda}\right) \sqrt{-g} \\
     \frac{\p \Lc_e}{\p(\p_\lambda e_\alpha^{\,\,\,\mu})} 
                  & = & -\frac{\hbar c}{4}\overline{\psi} 
          \gamma_5 \hat{\gamma}_\delta \psi \,\epsilon^{\delta\alpha\beta\gamma} 
                e_\gamma^{\,\,\,\lambda} e_{\beta\mu} \sqrt{-g}
\end{eqnarray}

\noindent
which give

\begin{eqnarray}
       T_{\mu\nu} &=& \frac{i}{4} \hbar c \bigl[\overline{\psi} \gamma_\mu \p_\nu \psi 
  + \overline{\psi} \gamma_\nu \p_\mu \psi - (\p_\mu \overline{\psi}) \gamma_\nu \psi 
  - (\p_\nu \overline{\psi}) \gamma_\mu \psi \bigr]  \no \\
  & &   - \frac{1}{2} \hbar c q \left(\overline{\psi} \gamma_\mu \psi A_\nu 
               + \overline{\psi} \gamma_\nu \psi A_\mu \right)  \no \\
  & & + \frac{\hbar c}{4} \overline{\psi} \gamma_5 \hat{\gamma}_\delta \psi 
  \,\epsilon^{\delta\alpha\beta\gamma} e_\gamma^{\,\,\,\lambda} \bigl[ (e_{\alpha\mu} 
  \p_\lambda e_{\beta\nu} + e_{\alpha\nu} \p_\lambda e_{\beta\mu}) \no  \\
  & &      - \frac{1}{2} (e_{\alpha\mu}  \p_\nu e_{\beta\lambda} 
                + e_{\alpha\nu} \p_\mu e_{\beta\lambda} )\bigr]
\end{eqnarray}

The Lagrange form of the gravitational field equations
can now be expressed in terms of tetrads.
We first substitute $ g^{\mu\nu}   = \eta^{\alpha\beta} \, 
 e_\alpha^{\,\,\,\mu} e_\beta^{\,\,\,\nu} $ in $ \Lc_g $ to find [2]

\beq
      \frac{\p \Lc_g}{\p g^{\mu\nu} }
      - \p_\lambda \frac{\p \Lc_g}{\p (\p_\lambda g^{\mu\nu} )} =
         \frac{1}{2} e_{\alpha\nu} \biggl[ \frac{\p \Lc_g}{\p e_\alpha^{\,\,\,\mu}}
        - \p_\lambda \frac{\p \Lc_g}{\p (\p_\lambda e_\alpha^{\,\,\,\mu})} \biggr]
\eeq
This, together with the energy tensor for the electron,

\beq
  T_{\mu\nu} =    e_{\alpha\nu} 
     \biggl[ \frac{\p \Lc_e}{ \p e_\alpha^{\,\,\,\mu}} 
                   - \p_\lambda \frac{\p \Lc_e}{\p(\p_\lambda 
                      e_\alpha^{\,\,\mu})} \biggr]
\eeq
give the field equations 

\beq
        \frac{\p \Lc}{ \p e_\alpha^{\,\,\,\mu}} 
                   - \p_\lambda \frac{\p \Lc}{\p(\p_\lambda 
                     e_\alpha^{\,\,\mu})} = 0
\eeq

\clearpage

\section*{\large {\bf Appendix C: Energy Conservation.}}

The principle of energy conservation derives from the Lagrange equations of motion, by means
of the Hamilton function

\beq
        \H = \sum_{\phi}  c \pi \, \p_0 \phi - \Lc
\eeq
The fields $ \phi $ include $ e_\alpha^{\,\,\,\mu} , \;\psi, \;\overline{\psi}, \mbox{\rm and}\;
 A_\mu $, while the momenta are defined by

\beq
    c \pi = \frac{\p \Lc}{\p (\p_0 \phi)}
\eeq
The temporal change of $\H$ is

\begin{eqnarray}
   \p_0 \H  & = & \sum_{\phi} \p_0 \biggl[\frac{\p \Lc}{\p (\p_0 \phi)}\biggr] \p_0 \phi
          + \frac{\p \Lc}{\p (\p_0 \phi)} \p_0\p_0 \phi - \frac{\p \Lc}{\p \phi} \p_0 \phi
                - \frac{\p \Lc}{\p (\p_\lambda \phi)} \p_0 \p_\lambda \phi     \no \\
         & = & \sum_{\phi} \p_0 \biggl[\frac{\p \Lc}{\p (\p_0 \phi)}\biggr] \p_0 \phi
       - \frac{\p \Lc}{\p \phi} \p_0 \phi - \frac{\p \Lc}{\p (\p_n \phi)} \p_0 \p_n \phi
\end{eqnarray}
Making use of

\beq
     \p_n \biggl[\frac{\p \Lc}{\p(\p_n \phi)} \p_0 \phi \biggr] =
                \p_n \biggl[\frac{\p \Lc}{\p(\p_n \phi)} \biggr] \p_0 \phi
                                    + \frac{\p \Lc}{\p(\p_n \phi)} \p_n \p_0 \phi
\eeq
we find

\begin{eqnarray}
      \p_0 \H  & = & \sum_{\phi} \biggl[\p_\lambda \frac{\p\Lc}{\p(\p_\lambda \phi)}
                             - \frac{\p \Lc}{\p \phi} \biggr] \p_0 \phi
                     - \p_n\biggl[\frac{\p\Lc}{\p(\p_n \phi)}\p_0\phi\biggr]     \no\\
        & = & \sum_{\phi}  - \p_n\biggl[\frac{\p\Lc}{\p(\p_n \phi)}\p_0\phi\biggr]
\end{eqnarray}
The last step is by virtue of the Lagrange equations of motion.  Integrate over all
3-dimensional space and discard surface terms, in order to obtain conservation of total energy

\beq
   \frac{d}{dx^0} \int \H \, d^3 x = 0
\eeq

\indent
We now derive an explicit expression for the energy density

\beq
\H = c\pi^\alpha_{\,\,\,\mu} \,\p_0 e_\alpha^{\,\,\,\mu} + c\overline{\pi} \,\p_0 \psi
        + \p_0\overline{\psi} \, c\pi + c\pi^\mu \,\p_0 A_\mu - \Lc
\eeq

\noindent
The spinor momenta are

\begin{eqnarray}
    c\overline{\pi} & = & \frac{\p\Lc}{\p(\p_0\psi)} =
           \frac{i}{2}\hbar c\sqrt{-g} \, \overline{\psi}\,\gamma^0  \\
  c\pi & = & \frac{\p\Lc}{\p(\p_0\overline{\psi})} =
                            - \frac{i}{2} \hbar c \sqrt{-g} \, \gamma^0 \psi
\end{eqnarray}
The electromagnetic momenta follow from

\beq
 L_{e-m} = - \frac{1}{4} \left(2 g^{00}g^{ij} F_{i0}F_{j0} + g^{il}g^{jm} F_{ij}F_{lm} \right)
\eeq

\beq
             \frac{\p F_{j0}}{\p(\p_0 A_i)} = - \delta^i_j
\eeq
Therefore,

\begin{eqnarray}
       c\pi^0 & = & \frac{\p \Lc}{\p(\p_0 A_0)} = 0          \\
      c \pi^i & = & \frac{\p \Lc}{\p(\p_0 A_i)}  = \sqrt{-g}\, F^{i0}
\end{eqnarray}

\indent
Turning now to the tetrad momenta, the gravitational Lagrangian

\begin{eqnarray}
   L_g & = & \frac{\kappa}{2} \, g^{\mu\nu}
             \left(\Gamma^\lambda_{\mu\nu} \Gamma^\rho_{\rho\lambda}
                - \Gamma^\lambda_{\rho\nu}\Gamma^\rho_{\mu\lambda}\right)          \no \\
       & = & \frac{\kappa}{2}
       \biggl\{ g^{00}\left(\Gamma^l_{m0}\Gamma^m_{l0}  - \Gamma^l_{l0}\Gamma^m_{m0}\right)
         + g^{lm}\left(\Gamma^n_{lm}\Gamma^0_{0n} - \Gamma^0_{0l}\Gamma^p_{pm}\right)  \no \\
       & & \hspace{.5in}  + g^{lm}\left(\Gamma^n_{lm}\Gamma^p_{pn}
                - \Gamma^p_{nl}\Gamma^n_{mp}\right) \biggr\} 
\end{eqnarray}
contains time derivatives in the first term

\beq
     \frac{\p \Gamma^l_{m0}}{\p(\p_0 g^{ij})} = -\frac{1}{2} g_{im} \delta^l_j \hspace{.3in}
      \frac{\p \Gamma^l_{l0}}{\p(\p_0 g^{ij})} = -\frac{1}{2} g_{ij} 
\eeq

\noindent
Therefore,

\beq
  \frac{\p \Lc_g}{\p(\p_0 g^{ij})} = -\frac{\kappa}{2} \sqrt{-g} \, g^{00} 
     \left(g_{il} \Gamma^l_{j0} - g_{ij} \Gamma^l_{l0} \right)
\eeq
\newpage 
\noindent
The electron Lagrangian (28) contains time derivatives in the coupling term

\begin{eqnarray}
   \frac{\p \Lc_e}{\p(\p_0 e_\beta^{\,\,\,\mu})} 
    & = & \frac{\hbar c}{4} \overline{\psi} \gamma_5 \hat{\gamma}_\delta \psi 
  \,\epsilon^{\delta\alpha\beta\gamma} e_\gamma^{\,\,\,0} e_{\alpha\mu} \sqrt{-g} \no \\
    & = & \frac{\hbar c}{4} \overline{\psi} \gamma_5 \hat{\gamma}_\delta \psi 
  \,\epsilon^{\delta\alpha\beta 0} e_0^{\,\,\,0} e_{\alpha\mu} \sqrt{-g}
\end{eqnarray}
which leaves

\beq
    \frac{\p \Lc_e}{\p(\p_0 e_a^{\,\,\,i})} 
     = - \frac{\hbar c}{4} \overline{\psi} \gamma_5 \hat{\gamma}_d \psi 
       \,\epsilon^{dab0} e_0^{\,\,\,0} e_{bi} \sqrt{-g}
\eeq
The tetrad momenta are given by

\begin{eqnarray}
  c\pi^\alpha_{\,\,\,\mu} & = & \frac{\p \Lc}{\p(\p_0 e_\alpha^{\,\,\,\mu})}   \no  \\
     &=& \frac{\p \Lc_g}{\p (\p_0 g^{\nu\lambda})} 
                         \frac{\p(\p_0 g^{\nu\lambda})}{\p(\p_0 e_\alpha^{\,\,\,\mu})} 
                           + \frac{\p \Lc_e}{\p(\p_0 e_\alpha^{\,\,\,\mu})}  \no  \\
      &=& \frac{\p \Lc_g}{\p (\p_0 g^{\nu\lambda})} 
       \left(\delta^\nu_\mu e^{\alpha\lambda} + \delta^\lambda_\mu e^{\alpha\nu} \right) 
          + \frac{\p \Lc_e}{\p(\p_0 e_\alpha^{\,\,\,\mu})}   
\end{eqnarray}
It follows that

\begin{eqnarray}
     c \pi^0_{\,\,\,0} & = & \frac{\p \Lc}{\p(\p_0 e_0^{\,\,\,0})} = 0                \\
     c \pi^a_{\,\,\,i}  & = & \frac{\p\Lc}{\p(\p_0 e_a^{\,\,\,i})}                  \no \\
                              & = & -\kappa \sqrt{-g}\, g^{00} 
            \left(e^a_l \Gamma^l_{i0} - e^a_{\,\,\,i} \Gamma^l_{l0} \right)
         - \frac{\hbar c}{4} \overline{\psi} \gamma_5 \hat{\gamma}_d \psi 
                                  \, \epsilon^{dab0} e_0^{\,\,\,0} e_{bi} \sqrt{-g}  \no   \\
\end{eqnarray}

\clearpage
\noindent
Substituting the momenta into (138), we find the total energy density to be

\begin{eqnarray}
\H & = & \frac{\kappa}{2} 
         \sqrt{-g} \biggl\{ g^{00} \left( \Gamma^l_{m0}\Gamma^m_{l0} 
               - \Gamma^l_{l0}\Gamma^m_{m0} \right) 
             - g^{lm}\left( \Gamma^n_{lm}\Gamma^0_{0n} 
                   - \Gamma^0_{0l}\Gamma^p_{pm}\right)                                 \no \\
& & \hspace{.5in} - g^{lm}\left( \Gamma^n_{lm}\Gamma^p_{pn} 
                   - \Gamma^p_{nl}\Gamma^n_{mp} \right) \biggr\}                        \no \\
& &     + \frac{1}{2} \sqrt{-g} \, g^{00} g^{lm} \left(\p_l A_0 \,\p_m A_0 
    - \p_0 A_l \, \p_0 A_m\right) + \frac{1}{4} \sqrt{-g}\, g^{il} g^{jm} F_{ij} F_{lm}  \no \\
& &     - \frac{i}{2}\hbar c\bigl[\overline{\psi} \gamma^l \p_l \psi 
        - (\p_l \overline{\psi})\gamma^l \psi \bigr]  \sqrt{-g}                             
    + \hbar c q \overline{\psi} \gamma^\mu \psi A_\mu \sqrt{-g}                       \no \\ 
& &  \hspace{.5in}  - \frac{\hbar c}{4} \overline{\psi} \gamma_5 \hat{\gamma}_0 \psi
          \,\epsilon^{0abc} e_c^{\,\,\,l} e_{am} \p_l e_b^{\,\,\,m} \sqrt{-g}                  
\end{eqnarray}

\clearpage

\section*{\large {\bf  References.}}

\begin{enumerate}

\item K. Dalton, ``Electromagnetism and Gravitation,'' {\it Hadronic Journal}
 {\bf 17} (1994) 483; www.arxiv.org/gr-qc/9512027.
\item K. Dalton, ``Gravity and the Dirac Electron,'' 
   \newline www.arxiv.org/physics/0409042.
\item L.D. Landau and E.M. Lifshitz, {\it The Classical Theory of Fields} 
  \newline (Pergamon, 1975).
\item T-P Cheng and L-F Li, {\it Gauge Theory of Elementary Particle Physics}
 (Oxford, 1984).
\item M. Guidry, {\it Gauge Field Theories} (Wiley, 1991).
\item W. Rolnick, {\it The Fundamental Particles and Their Interactions}
  \newline (Addison-Wesley, 1994).

\item S. Weinberg, {\it Gravitation and Cosmology} (Wiley, 1972).
\item V. de Sabbata and M. Gasperini, {\it Introduction to Gravitation} 
  \newline (World Scientific, 1985).

\end{enumerate}

\end{document}